\documentclass{WileyMSP-template}
\usepackage{color}

\begin{document}

\pagestyle{fancy}
\rhead{\includegraphics[width=2.5cm]{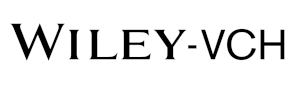}}
.

\title{Variable Dual Auxeticity of the Hierarchical Mechanical Metamaterial composed of Re-entrant Structural Motifs}

\maketitle

% Author: Please give full first and last names for authors and include * after the name of all corresponding authors

\author{Krzysztof K. Dudek*}
\author{Julio A. Iglesias Martínez}
\author{Muamer Kadic}

% Dedication

%\dedication{Optional dedication here. If no dedication is required, please leave blank}

% Affiliations: Please provide adacemic titles (Prof. or Dr.) for all authors where applicable, and include an institutional email address for all corresponding authors
\begin{affiliations}
K. K. Dudek\\
Address: Institute of Physics, University of Zielona Gora, ul. Szafrana 4a, Zielona Gora, 65-069 Poland\\
Email Address: k.dudek@if.uz.zgora.pl

K. K. Dudek, J. A. Iglesias Martínez, M. Kadic\\
Address: Institut FEMTO-ST, CNRS, Université Bourgogne Franche-Comté, Besançon, 25030 France

\end{affiliations}

% Keywords: Please provide a minimum of three and a maximum of seven keywords, separated by commas


% Abstract should be written in the present tense and impersonal style (i.e., avoid we), and be at most 200 words long
\begin{abstract}

In this work, a novel hierarchical mechanical metamaterial is proposed that is composed of re-entrant truss-lattice elements. It is shown that this system can deform very differently and can exhibit a versatile extent of the auxetic behaviour depending on a small change in the thickness of its hinges. In addition, depending on which hierarchical level is deforming, the whole structure can exhibit a different type of auxetic behaviour that corresponds to a unique deformation mechanism. This results in a dual auxetic structure where the interplay between the two auxetic mechanisms determines the evolution of the system. It is also shown that depending on the specific deformation pattern, it is possible to observe a very different behaviour of the structure in terms of frequencies of waves that can be transmitted through the system. In fact, it is demonstrated that even a very small change in the parametric design of the system may result in a significantly different band gap formation that can be useful in the design of tunable vibration dampers or sensors. The possibility of controlling the extent of the auxeticity also makes the proposed metamaterial to be very appealing from the point of view of protective and biomedical devices.

%In this work, a novel hierarchical mechanical metamaterial is proposed that is composed of re-entrant truss-lattice elements. It is shown that this system can deform very differently and can exhibit a different extent of the auxetic behaviour depending on a small change in the thickness of its hinges. In addition, depending on which hierarchical level is deforming, the whole structure can exhibit a different type of the auxetic behaviour that corresponds to a unique deformation mechanism. This results in a dual auxetic structure where the interplay between the two auxetic mechanisms determines the evolution of the system. It is also shown that depending on the specific deformation pattern, it is possible to observe a very different behaviour of the structure in terms of the frequencies of waves that can be transmitted through the system. In fact, it is demonstrated that even a very small change in the parametric design of the system may result in the significantly different band gap formation that can be useful in the design of tunable vibration dampers or sensors. The possibility of controlling the extent of the auxeticity also makes the proposed metamaterial to be very appealing from the point of view of protective and biomedical devices.   

\end{abstract}

% Text: Please use section headings and subheadings as specified below. For communications, all section headings apart from Experimental Section should be removed
% Please make the first reference to a display item bold: \textbf{Figure 1}
% Do not abbreviate Figure, Equation, etc.; display items are always singular, i.e., Figure 1 and 2.
% Equations are always singular, i.e., Equation 1 and 2, and should be inserted using the {equation} environment, not as graphics
% Please do not use footnotes in the text, additional information can be added to the Reference list.

\section{Introduction}

%Mechanical metamaterials -> example: Truss lattice mechanical metamaterials
Mechanical metamaterials \cite{Florijn2014, Coulais_Teomy_2016, Milton_expanders_2013, Coulais2018, christensen2015vibrant, Neville_2016, Jiang_Chen_2020, Frenzel_Wegener_2017, kadic20193d, Krushynska_2022, Mizzi_Int_J_Solid_2022, Mizzi_Mater_today_2020, Dudek_Marc_2021} are rationally-designed structures capable of exhibiting a plethora of atypical mechanical properties that are rarely observed in the case of naturally-occurring materials. Some of the most commonly studied of such properties are negative Poisson's ratio \cite{Almgren1985, Wojciechowski_1989, Wojciechowski_Branka_1989, Lakes1987, Evans_Alderson_2000, Babaee_2013, Lim_2019_compos, Novak_Biasetto_2021, Jiang_Ren_2022, Grima_squares_2000, Novak_Duncan_2021, Narojczyk_pssb_2022, Narojczyk_Bilski_2022, Dunncan_SMS_2022, Tretiakov_pssb_2020, Hoover2005} (auxetic behaviour), negative stiffness \cite{Bertoldi_Vitelli_2017, Hewage2016, Chen_Tan_2021, Dudek_Roy_Soc_2018, Tan_Wang_2022} and negative compressibility \cite{Lakes_Wojciechowski_2008, Dudek_triangles_2016}. Over the last thirty years, it has been demonstrated that devices utilising materials exhibiting such characteristics can be used in the case of a variety of applications including protective devices \cite{Imbalzano2018, Miniaci2016}, sports equipment \cite{Duncan_sport_2018} as well as biomedical \cite{Kolken_Zadpoor2018, Teunis_van_Manen_2021} and vibration damping devices \cite{Quadrelli2021, Guenneau_Ramakrishna_2007, Fleury2014}. One of the most studied of these unusual mechanical properties seems to be auxetic behaviour where the keen interest of researchers in this property stems from the fact that auxetic materials often exhibit high indentation resistance \cite{Li_Liu_2020}, wave attenuation \cite{Ruzzene_2005, Qi_Yu_2019} and many other features that are useful in the case of various applications. The commercial appeal of auxetic mechanical metamaterials resulted in a broad range of studies focused on different types of mechanical structures. One of the most commonly investigated classes of such systems are truss-lattice-based auxetic mechanical metamaterials. \vspace{4 mm}

% Truss lattice mechanical metamaterials: lightweight, easy additive manufacturing etc. -> Limitations -> Hierarchical mechanical matamaterials
In recent years, auxetic mechanical metamaterials in the form of truss-lattice structures have been a subject of numerous studies \cite{Mizzi_truss_lattice, Larsen_1997, Dudek_arrowhead_3D, Guo_Yang_2022, Lim_arrowhead, Dudek_mater_des_2020} with a particular emphasis on the design of efficient protective materials. This stems from the fact that in addition to their ability to exhibit the negative Poisson's ratio, such systems are also very light in comparison to many other auxetic mechanical metamaterials. This, in turn, is of great significance when designing lightweight protective devices that can exhibit significant energy absorption similarly to their heavier counterparts while at the same time they do not significantly increase the mass of the vehicle/object that they protect. One of the earliest and most famous examples of such structures corresponds to the so-called arrow-head mechanical metamaterial that can be used both as a 2D \cite{Larsen_1997} and 3D \cite{Dudek_arrowhead_3D, Guo_Yang_2022, Lim_arrowhead, Dudek_mater_des_2020} system characterised by the negative Poisson's ratio. Thanks to the success of this and other similar re-entrant mechanical metamaterials \cite{Yang_2015, Liu_Wang_2016, Dudek_Kadic_2021, Fu_Xu_2016, Tan_Small_2022}, over the years, it has been possible to observe a large number of truss-lattice auxetic mechanical metamaterials that are typically working based on a similar hinging mechanism of mutually-connected truss elements. Nevertheless, despite the numerous advantages of standard truss-lattice auxetic mechanical metamaterials, they normally share several limitations. Most importantly, such structures typically correspond to a specific set of mechanical properties that cannot be significantly altered once the structure is manufactured. In fact, a similar limitation often applies to a vast majority of other known mechanical metamaterials. However, in recent years, researchers reported a few approaches that allow to overcome this shortcoming. It seems that one of the most promising approaches allowing to observe numerous different mechanical properties and deformation patterns, without the need of significantly changing or rebuilding the system, corresponds to hierarchical mechanical metamaterials. \vspace{4 mm}

Hierarchical mechanical metamaterials \cite{Lakes_hierarchical_1993, Tang_Lin_hierarchical_2015, Gatt_hierarchical_2015, An_Bertoldi_2020, Oftadeh_2014, Dudek_Materials_hierarchical2021, Cho_hierarchical_2014, Dudek_2022_Adv_Mater} are a class of structures composed of multiple structural levels having their own geometry that typically can deform irrespective of each other. This, in some cases, allows the system to exhibit quantitatively very different mechanical properties without the need of being reconstructed. This, in turn, significantly enhances the applicability of hierarchical structures. In principle, the concept of hierarchical mechanical metamaterials is not limited to a specific type of mechanical metamaterials and can be also applied in the case of truss-lattice systems. In the literature, there are several examples of such structures that are based on re-entrant honeycombs \cite{Oftadeh_2014} as well as other structural motifs \cite{Mizzi_truss_lattice}. Nevertheless, this direction of studies is still in its infancy and many new studies have to be conducted in order to better utilise the potential of these structures as well as to fully understand the underlying physics. One of such relatively unexplored aspects related to this class of systems corresponds to the possibility of controlling the band gap formation \cite{Cho_hierarchical_2014, Kunin2016} that would allow to construct tunable vibration insulators or sensors for different ranges of frequencies. Another aspect corresponds to the control over the mechanical properties (e.g. auxeticity) of the hierarchical structure. In this case, despite the success of several projects and their commercial appeal, there are still relatively few known efficient hierarchical mechanisms in comparison to standard non-hierarchical mechanical metamaterials. \vspace{4 mm}

In this work, a novel hierarchical mechanical metamaterial composed of truss-lattice elements is proposed and analysed from the point of view of its mechanical properties. More specifically, it is demonstrated that depending on the thickness of different groups of its hinges, the considered structure can follow different deformation patterns. In addition, depending on which hierarchical level is deforming, the whole structure can exhibit a different type of auxetic behaviour that corresponds to a unique deformation mechanism. This results in a dual auxetic structure where the interplay between the two auxetic mechanisms determines the evolution of the system. Finally, it is also presented that depending on which hierarchical level of the structure is deforming, one can observe a drastically different phonon dispersion of the system that also leads to a very different band gap formation. 

% Hierarchical metamaterials advantages

%This work

\section{Model}

\subsection{Design}
\begin{figure}
	\centering
	\includegraphics[width=0.9\linewidth]{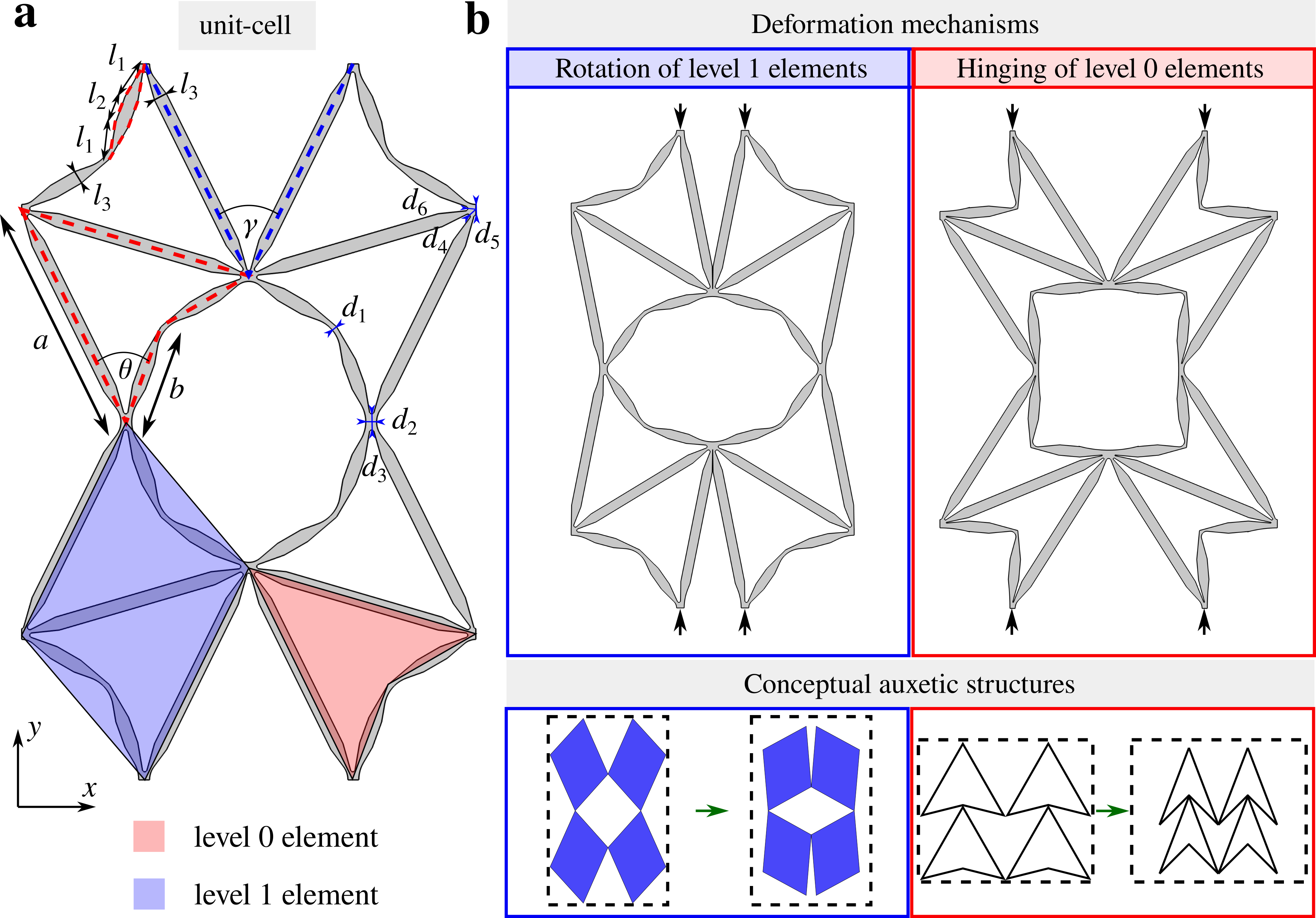}
	\caption{The unit-cell of the structure analysed in this work from the point of view of its mechanical properties. {\bf a)} A diagram of the unit-cell of the system with the definition of all of the dimensions and geometric parameters of the structure. {\bf b)} Different deformation patterns exhibited by the considered system in the case of the isolated deformation of the level 0 or level 1 elements of the hierarchical structure.\label{model}}
\end{figure}

The main objective of this work is to design a hierarchical mechanical metamaterial capable of exhibiting versatile magnitudes of the negative Poisson's ratio that can be adjusted by a small variation in the design parameters with an emphasis on the thickness of hinges. To this aim, the considered model corresponds to the two-dimensional hierarchical mechanical metamaterial composed of truss-lattice elements as shown in Fig. \ref{model}. \vspace{4 mm}

The considered system is a two-level hierarchical structure composed of truss-lattice elements having the thickness of $l_{3}$ (see Fig. \ref{model}(a)). Level 1 of the structure corresponds to parallelogram-like elements connected to each other at vertices similarly to the well-known rotating squares \cite{Grima_squares_2000} or rotating rhombi \cite{Grima_Farrugia_2008} systems that had been reported to be able to exhibit auxetic behaviour. On the other hand, level 0 elements assume the form of two arrow-head units that share one edge having a length denoted as $a$. It is important to remember that as mentioned in the Introduction, mechanical metamaterials consisting of arrow-head-based structural elements are known in the literature and, as shown in diagrams presented in Fig. \ref{model}(b), can exhibit a strong auxetic behaviour. This means that both level 0 and level 1 of the considered hierarchical system correspond to very different deformation mechanisms that can lead to the entire system exhibiting a negative Poisson's ratio. For the considered model, such observation is of great significance since the two hierarchical levels can deform independently. This means that level 1 blocks can rotate with respect to each other while retaining their shape. Thus, such a process can occur without hinging of truss-lattice beams corresponding to level 0 elements. Conversely, it is possible to observe hinging of level 0 elements without the rotation of level 1 blocks. In the case of this model, the hinging of level 0 elements is quantified by a change in the value of $\theta$ (see Fig. \ref{model}(a)). On the other hand, the rotation of level 1 blocks is quantified by a variation in the value of $\gamma$. As a result, once the structure is manufactured and specific geometric dimensions are well-defined, the magnitudes of angles $\theta$ and $\gamma$ provide full information about the configuration assumed by the structure assuming no flexure of truss-lattice elements.   \vspace{4 mm}

The factor that determines the extent of the deformation of level 0 and level 1 of the structure is the relative thickness of hinges connecting structural elements corresponding to different hierarchical levels. These hinges are denoted as $d_{1}$-$d_{5}$ in Fig. \ref{model}(a) and in general can assume very different values. As demonstrated schematically in Fig. \ref{model}(b), it is expected that upon changing the relative thickness of hinges connecting truss-beams corresponding to level 0 and level 1 elements, it should be possible to increase the extent of deformation of one hierarchical level over another. In the extreme case, if the difference in the thickness of the two groups of hinges was very considerable, it could be even possible to observe an isolated deformation either of level 0 or level 1 of the system. Such possibility is shown in Fig. \ref{model}(b) where in the first scenario one can see the deformation of the structure corresponding solely to the rotation of level 1 blocks while in the second case the deformation corresponds exclusively to hinging of level 0 elements. However, in many situations, it can be expected that the mixture of the two deformation mechanisms could be observed.  \vspace{4 mm}

At this point, it should be also noted that the two types of truss-lattice beams present within the system have a very specific shape. Namely, both ends of such beams having the length of either $a$ or $b$ (see Fig. \ref{model}(a)) are significantly narrower than their central parts. This particular aspect of the design of the considered system allows ensuring that its deformation will occur primarily via the relative hinging of truss-lattice beams and that their flexure will be negligible.

\subsection{Simulations}

To assess the mechanical properties of the considered system, all models are deformed via the compression along the $y$-axis as schematically indicated by means of black arrows in Fig. \ref{model}(b). Furthermore, since only a single unit-cell is taken into account, in order to describe the behaviour of the system, it is assumed that periodic boundary conditions are imposed \cite{Mizzi_PBC} on the left and right edge of the unit-cell. These boundary conditions ensure that the two edges retain the form of a straight line throughout the deformation process. In order to simulate the behaviour of the structure, Finite Element Method (FEM) simulations were conducted through the use of the COMSOL Multiphysics software. In order to obtain realistic results, the nonlinear geometry feature was enabled while it was assumed that the material itself was isotropic. More specifically, the respective properties of the material used for the considered model were set to be the following for all conducted simulations. Youngs's modulus: 4 GPa, Poisson's ratio: 0.4 and density: 1200 kg m$^{-3}$. These material properties correspond to some of the resins \cite{Fehima_2022} that could be used in order to 3D print the considered structure. The remaining geometric parameters used in the simulations were set to be the following: $a$ = 3.5 cm, $b$ = 1.5 cm, $\theta (t = 0)$ = 46.24$^{\circ}$ (initial value), $\gamma (t = 0)$ = 10$^{\circ}$ (initial value), $l_{1}$ = 5.5 mm, $l_{2}$ = 4 mm, $l_{3}$ = 2 mm.

\section{Results and Discussion}

In order to assess the ability of the considered system to exhibit versatile deformation patterns and different values of the negative Poisson's ratio, three different configurations of the considered model were selected. These configurations named case 1, case 2 and case 3 are almost identical with the only difference between them being the thickness of hinges $d_{1}$ - $d_{6}$. Specific values of these parameters are provided in Table \ref{table_param}.

\begin{table}[]
	\centering
	\begin{tabular}{|l|c|c|c|c|c|c|}
		\hline
		& $d_{1}$ [mm] & $d_{2}$ [mm] & $d_{3}$ [mm] & $d_{4}$ [mm] & $d_{5}$ [mm] & $d_{6}$ [mm] \\ \hline
		case 1 & 1.6 & 0.6 & 8.0 & 4.0 & 0.6 & 4.0\\ \hline
		case 2 & 0.2 & 2.5 & 0.4 & 0.2 & 0.4 & 0.2\\ \hline
		case 3 & 0.8 & 1.6 & 1.6 & 1.4 & 1.6 & 0.8 \\ \hline
	\end{tabular}
	\caption{The thickness of hinges corresponding to three types of configurations of the considered model.  \label{table_param}}
\end{table}

\subsection{Controllable auxetic behaviour}

\begin{figure}
	\centering
	\includegraphics[width=0.8\linewidth]{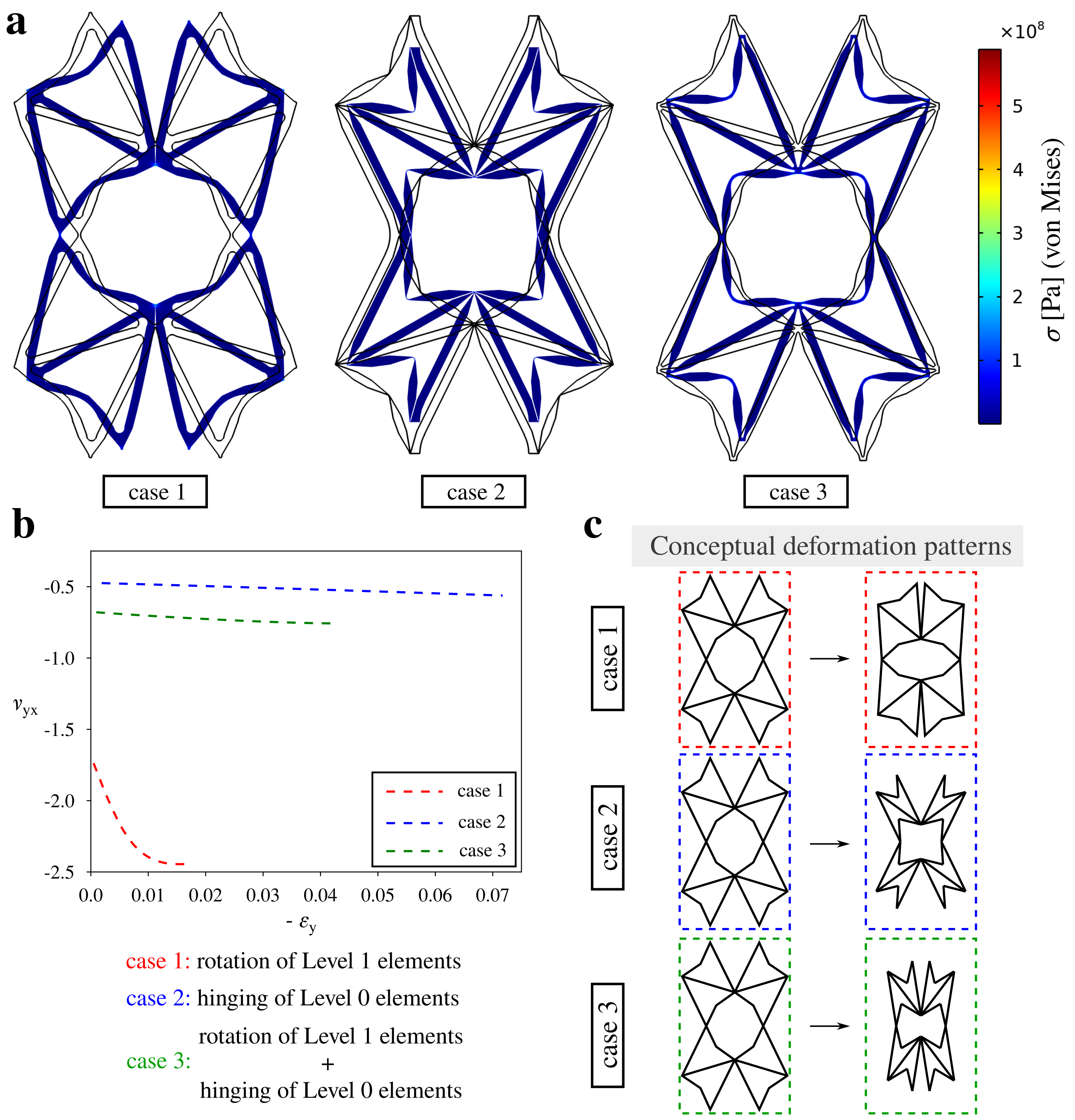}
	\caption{Deformation patterns and Poisson's ratio of the three versions of the considered model. {\bf a)} Deformation of the three analysed cases subjected to the compression along the $y$-axis. For each of the examples, the black outline in the background corresponds to the initial shape of the structure. The extent of mechanical deformation is graphically enhanced by a factor of 2 to visually emphasise the difference between different deformation mechanisms. The deformation processes corresponding to structures named case 1, case 2 and case 3 are also presented as animations provided in the form of the {\bf Supplementary Video 1}, {\bf Supplementary Video 2} and {\bf Supplementary Video 3}. {\bf b)} Engineering Poisson's ratio for all types of the system. {\bf c)} Conceptual diagrams showing differences between the three types of possible deformation patterns.   \label{results_main}}
	
\end{figure}

As shown in Fig. \ref{results_main}, the three different versions of the considered system exhibit very distinct deformation patterns. According to Fig. \ref{results_main}(a), the structure named case 1 deforms primarily via the rotation of level 1 blocks while level 0 elements do not undergo the hinging process. This stems from the fact that as specified in Table \ref{table_param}, hinges within the arrow-head-shaped structural elements (in particular the hinge having its thickness defined as $d_{1}$) are in this case very thick in comparison to hinges connecting level 1 blocks (primarily hinges having their thickness defined as $d_{2}$ and $d_{5}$ although all hinges with the exception for the hinge having the thickness of $d_{1}$ affect the rotation of level 1 elements). A very different behaviour can be observed in the case of the structure named case 2 where level 1 blocks seemingly do not rotate while level 0 elements undergo a hinging process to a significant extent. This very different deformation pattern can be explained by the fact that in this scenario, some of the hinges connecting level 1 blocks are very thick. On the other hand, hinges connecting truss beams within arrow-head structural elements are very thin. Finally, it is also important to note that as demonstrated in the case of the deformation of the structure referred to as case 3, it is possible to observe the simultaneous rotation of level 1 elements and hinging of level 0 structural motifs. For better clarity, all three of these deformation mechanisms are schematically shown in Fig. \ref{results_main}(c). At this point, it is also worth mentioning that in the hypothetical scenario where all of the hinges present within the system would have the same thickness, the behaviour of the structure would resemble that of the system referred as case 3 (see Supplementary Information).\vspace{4 mm}

\begin{figure}
	\centering
	\includegraphics[width=0.8\linewidth]{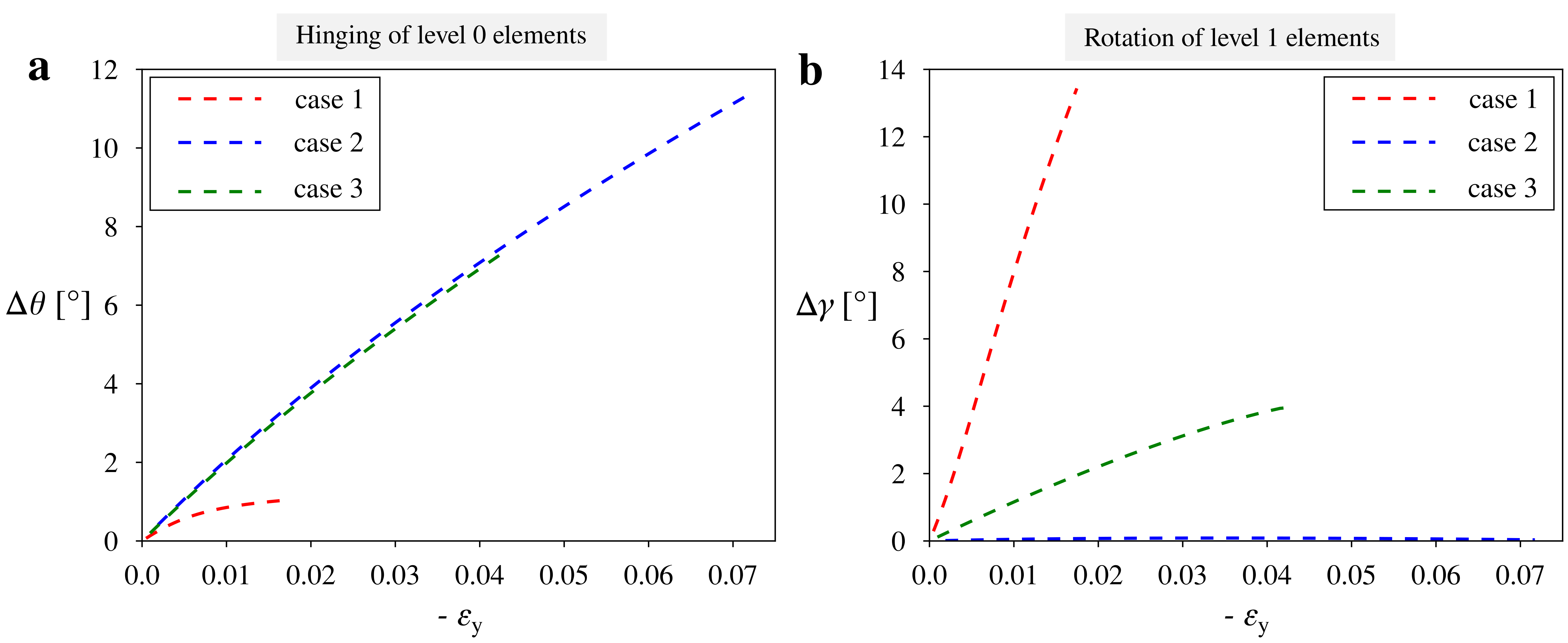}
	\caption{Variation in the angles describing the deformation of level 0 and level 1 of the considered hierarchical structure.}
	\label{results_graphs}
\end{figure}

In addition to the possibility of conducting the qualitative comparison of the bahaviour of the three types of the considered model, it is also possible to conduct the quantitative analysis where the variation in specific geometric parameters would be analysed. As shown in Fig. \ref{results_graphs}, the above qualitative discussion matches the variation in angles $\theta$ and $\gamma$ that quantify the deformation of level 0 and level 1 respectively. More specifically, one can note that angle $\theta$ changes significantly for case 2 and case 3 structures (see Fig. \ref{results_graphs}(a)) where both of these cases correspond to a significant hinging of level 0 elements. On the other hand, for the structure named case 1, it is possible to see a very small change in this parameter. Conversely, as presented in Fig. \ref{results_graphs}(b), the opposite trend can be seen in the case of the variation in the angle $\gamma$ that defines the extent of rotation of level 1 elements.  \vspace{4 mm}

As described above, the change in the thickness of some of the hinges can significantly change the deformation patterns exhibited by the structure during the compression. However, from the point of view of potential applications, it is essential to determine the ability of the considered structure to exhibit a negative Poisson's ratio. As shown in Fig. \ref{results_main}(b), all of the considered versions of the analysed system exhibit auxetic behaviour throughout the deformation process. This in turn is very important since it shows that a change in the deformation pattern does not affect the ability of the structure to exhibit a negative Poisson's ratio. Nonetheless, it does not mean that its magnitude does not change. According to Fig. \ref{results_main}(b), the Poisson's ratio corresponding to the case 1 structure assumes very low values that reach the magnitude of approximately -2.5. On the other hand, a significantly less negative value of the Poisson's ratio can be observed in the case of the deformation of the system referred to as case 2. For this structure, the magnitude of the Poisson's ratio initially assumes the value of -0.5 and becomes gradually lower as the vertical compression of the system continues. Furthermore, in the last of the analysed scenarios, the structure named case 3 exhibits the intermediate range of the negative Poisson's ratio with its values changing at the interval approximately between -0.6 and -0.7. One can note that the Poisson's ratio, in this case, is more similar to case 2 than the case 1 system. This can be explained by the fact that for this type of structure, the hinging of level 0 elements is more dominant than the rotation of level 0 elements (see Fig. \ref{results_main}(a) and Fig. \ref{results_graphs}). However, this does not mean that the intermediate case cannot exhibit the Poisson's ratio that is more similar to the auxeticity exhibited by the case 1 system should the thicknesses of its hinges be appropriately adjusted.

\subsection{Vibrational properties}

In addition to the possibility of controlling the deformation pattern and the extent of the exhibited auxeticity, it is also interesting to check whether different types of the considered model allow observing some differences in terms of the band gap formation and frequencies of vibrations that can propagate through the system. To achieve it, FEM simulations were conducted by means of the COMSOL Multiphysics software. In the case of each of the structures, to generate the results, Floquet periodic boundary conditions were implemented in the $x$- and $y$-directions assuming two-dimensional unit-cells. All of the design parameters were set to be identical as in the case of mechanical testing.  \vspace{4 mm}

\begin{figure}
	\centering
	\includegraphics[width=0.95\linewidth]{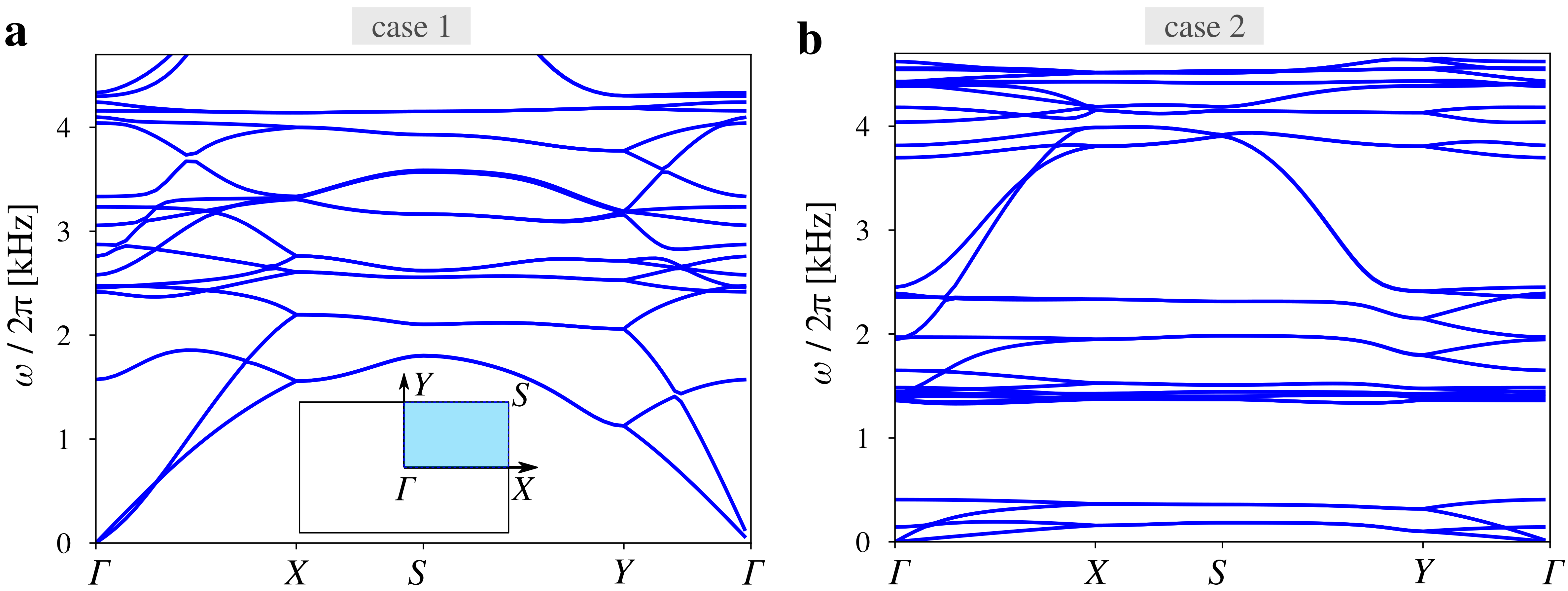}
	\caption{Phonon dispersion for the two types of structures considered in this work. The inset provided on panel (a) schematically depicts the first Brillouin zone corresponding to the considered model. Specific points presented at the centre as well as at the edge and corners of the first Brillouin zone indicate high-symmetry points that were used in the conducted analysis in order define directions in which the vibrational properties of the system were investigated.}
	\label{phonon_dispersion}
\end{figure}

Based on Fig. \ref{phonon_dispersion}, one can note that the phonon dispersion graphs corresponding to systems called case 1 and case 2 are very different. Most importantly, in the case of the first system, within the presented range of frequencies, there are no band gaps. On the other hand, the phonon dispersion associated with the case 2 structure corresponds to a band gap for frequencies approximately in the range between 0.5 kHz and 1.3 kHz. This, in turn, is very interesting as it indicates that a small change in the thickness of the hinges may cause the appearance of a considerable band gap where its size can be further modified by adjusting the relative thickness of the hinges. This means that the considered system can in general undergo a transition from the structure with no band gaps to the system working as a vibration damper where specific frequencies are not transmitted. Of course, in the current version of the proposed model such transition cannot occur in an active manner since the change in the thickness of hinges would require reconstructing the system or subjecting it to a post-manufacturing treatment that would alter some of its dimensions. However, this is not the only approach making it possible to change the behaviour of the considered system from the characteristic typical for the case 1 system to the behaviour of the case 2 structure and vice versa. This can be achieved for example by constructing hinges from another material corresponding to a very different thermal expansion coefficient than the rest of the system. Such a solution could allow changing the behaviour of the structure at different temperatures. Another approach could correspond to the use of electromagnets / magnets \cite{Dudek_mater_des_2020} embedded in specific parts of the unit-cell. The mutual interaction of such inclusions could adjust the effective stiffness of specific hinges within the system and as a result, would allow for the active transition of the structure. At this point, we would like to also emphasise the fact that according to the results presented in Fig. 4, the band gap formation can be controlled by the relative thickness of hinges within the system with similar results being observed for other hierarchical mechanical metamaterials known in the literature \cite{Kunin2016}. However, even though this topic does not belong to the scope of this work, the band gap formation could also be potentially influenced by the deformation of the system. Nevertheless, this aspect of the behaviour of the considered structure would have to be thoroughly analysed before reaching any conclusions.  \vspace{4 mm}

All of this is very important since, in this work, it is shown that the proposed model can exhibit versatile deformation patterns depending on the interplay between the two auxetic mechanisms present within the system. It is also shown that a modification in the thickness of different groups of hinges can significantly change the extent of the auxetic behaviour exhibited by the system. Both of these results are not commonly observed for standard mechanical metamaterials and can have multiple applications. More specifically, if one was to construct the active version of the considered system (e.g. by using mutually-interacting inclusions in the form of electromagnets to modify the effective stiffness of hinges), then it would be possible to significantly change auxeticity of the structure without the need of reconstructing the system. This would allow to construct highly-efficient protective devices that could adjust the extent of their auxeticity to a specific cause of the deformation. The concept presented in this work could be also very interesting from the point of view of biomedical devices. More specifically, if one was to construct the structure composed of multiple unit-cells corresponding to a different deformation pattern, then in general, it would be possible to arbitrarily modify the shape of the entire structure. If the mismatch in the Poisson's ratio between different parts of the system would be significant, it would be possible to observe considerable shape-morphing similarly to the concept proposed in \cite{Dudek_2022_Adv_Mater} (in the aforementioned study, it was demonstrated that an initial shape of the metamaterial subjected to the mechanical deformation can be significantly modified depending on the ratio in the Poisson's ratio associated with different parts of the structure). The possibility of exhibiting controllable shape-morphing could be used in the design of stents in order to better support specific parts of the blood vessel that are particularly weak. Finally, it is important to emphasise the fact that the results presented in this work are not limited to a specific scale. In fact, the proposed model can be even constructed at the microscale.

\section{Conclusion}
It is shown that the considered hierarchical structure can follow very different deformation patterns depending on the variation in the thickness of its hinges. This means that the proposed system can utilise very different deformation mechanisms which is very atypical and highly beneficial in comparison to typical mechanical metamaterials. Furthermore, in this work, it is presented that the considered model can exhibit a very broad range of the negative Poisson's ratio that can be controlled by adjusting the relative stiffness of hinges. This result, which originates from the interplay of two independent auxetic mechanisms present within the structure, indicates that in theory, the auxeticity of the system can be adjusted depending on the specific application. This, in turn, could prove to be very useful in the design of highly-efficient protective devices. Finally, it should be emphasised that different configurations of the considered system may result in a very distinct behaviour of the structure in terms of frequencies of waves that can be transmitted through the analysed metamaterial. In fact, it is demonstrated that even a very small change in the parametric design of the system may lead to a significantly different band gap formation that can be useful in the design of tunable vibration dampers or sensors.

% Experimental section

%\section{Experimental Section}
%\threesubsection{First part of experimental section}\\
%\threesubsection{Second part of experimental section}\\

%\medskip
%\textbf{Supporting Information} \par %Please delete the Suppporting Information statement if it is not applicable. Please supply Supporting Information in another file. Supporting information should not be provided in .tex format
%Supporting Information is available from the Wiley Online Library or from the author.

% Acknowledgements
\medskip
\textbf{Acknowledgements} \par %delete if not applicable))
K.K.D. acknowledges the support of the Polish National Science Centre (NCN) in the form of the grant awarded as a part of the SONATINA 5 program, project No. 2021/40/C/ST5/00007
under the name “Programmable magneto-mechanical metamaterials guided by the magnetic field”. 

This research was funded by the Polish Minister of Education and Science under the program “Regional Initiative of Excellence” in 2019–2023, project
No. 003/RID/2018/19, funding amount PLN 11 936 596.10.

% Statements
\medskip
\textbf{Conflict of Interest} \par %delete if not applicable))
The authors declare no conflict of interest.

\medskip
\textbf{Data Availability Statement} \par %delete if not applicable))
The data that support the findings of this study are available from the corresponding author upon reasonable request.

% References
\medskip

% Use the following code if you wish to generate your bibliography with BibTeX;
% replace the string "MSP-template" below with the name(s) of
% the BibTeX data base(s) you want to use.
% The resulting bibliography-output (the content of the .bbl file)
% must be pasted back into this file before submission.
% Please also include your BibTeX data base file(s) in your submission
% so that we can re-run BibTeX if necessary.
%
%\bibliographystyle{MSP}
\bibliography{manuscript_references_ver2}

\end{document}